# Recirculating Light Phase Modulator

**Haijin Huang[1], Xu Han[1,2], Armandas Balčytis[1], Aditya Dubey[1], Andreas Boes[1,3,4], Thach G. Nguyen[1], Guanghui Ren[1], Mengxi Tan[1], Yonghui Tian[2] and Arnan Mitchell[1]**

[1] Integrated Photonics and Applications Centre, School of Engineering, RMIT University, Melbourne, VIC 3001, Australia
[2]Key Laboratory for Magnetism and Magnetic Materials of MOE, School of Physical Science and Technology, Lanzhou University, Lanzhou,730000 Gansu, China
[3]School of Electrical and Electronic Engineering, The University of Adelaide, Adelaide, SA 5005, Australia
[4]Institute for Photonics and Advanced Sensing, The University of Adelaide, Adelaide, SA 5005, Australia

arnan.mitchell@rmit.edu.au

## Abstract

High efficiency and a compact footprint are desired properties for electro-optic modulators. In this paper, we propose, theoretically investigate and experimentally demonstrate a recirculating phase modulator, which increases the modulation efficiency by modulating the optical field several times in a non-resonant waveguide structure. The 'recycling' of light is achieved by looping the optical path that exits the phase modulator back and coupling it to a higher order waveguide mode, which then repeats its passage through the phase modulator. By looping the light back twice, we were able to demonstrate a recirculating phase modulator that requires nine times lower power to generate the same modulation index of a single pass phase modulator. This approach of modulation efficiency enhancement is promising for the design of advanced tunable electro optical frequency comb generators and other electro-optical devices with defined operational frequency bandwidths.

**Introduction**

Optical modulators are a vital component in communication systems [1] and microwave photonic devices [2-5] as they provide the means to translate high speed electrical signals into the optical domain, where they can be further processed or propagate over kilometer length scales in optical fibres while experiencing minimal losses. Modulators can also be used for achieving desired on-chip functionalities, such as non-reciprocal devices [6]. While many optical modulators have been demonstrated using materials, such as silicon [5], indium phosphide [7], 2D materials [8] and polymers [9], arguably one of the most attractive platforms used for electro-optical functionalities is lithium niobate, owing to its low optical losses and high-speed modulation capacity [1].

Lithium niobate modulators have been commercially available for several decades and have been crucial in the success of optical fibre communications that underpin the internet [1]. Traditional lithium niobate devices use titanium indiffused waveguides, however a new generation of thin film silicon and indium phosphide integrated photonics has emerged offering improved drive voltage, compactness and the potential for system integration [1]. More recently, a similar revolution has taken place in the field of lithium niobate photonics with the emergence of the thin-film lithium niobate waveguide platform offering high efficiencies for electro-optic interaction, and nonlinear optical processes due to the strong modal confinement [10-12]. Nevertheless, there are design choices that need to be considered for efficient modulators in this platform such as the trade-off between the length and bandwidth of modulators [13] or the trade-off for the closeness of the electrode to the waveguide [14] and microwave losses of the electrodes [12]. Optical structures have also been investigated to improve modulation efficiency. For example loop mirrors [15], and Fabry–Perot cavities [16] have been demonstrated to increase the electro-optical interaction length. Recently, a promising way to increase the interaction length between a waveguide and a thermal heating element was demonstrated, providing an enhanced efficiency for tuning the phase of light by 'recycling' or recirculating it serval times through the same multimode waveguide, achieved by looping the light back and coupling it to the multimode waveguide using mode multiplexers [17].

In this contribution, we investigate if a conceptually similar light recirculation approach can be used for high-speed electro-optical modulators and how this concept can help to increase the efficiency of phase modulators. We experimentally demonstrate that the recirculating phase modulator increases the modulation efficiency with the number of recirculations. This enabled us to reduce the power of the microwave signal by a factor of 9 when the light passed 3 times through the phase modulator, when compared to a standard single pass phase modulator with the same electrode dimensions to achieve the same modulation depth.

**Device Concept**

The efficiency of an electro-optic device is directly related to the interaction length between the electrode and the electro-optic waveguide. One option is to make this interaction region as long as possible. We explore an alternative where light is recirculated several times through a phase modulator to improve the modulation efficiency. This is fulfilled by looping the light through different orthogonal modes in a multimode waveguide and applying microwave modulation simultaneously to all the different copropagating modes. Fig. 1(a) illustrates the proposed recirculating modulator with two light recycling loops that is described as a case-in-point in detail as follows.

In the proposed recirculating modulator, the light is coupled into a single mode waveguide in the form of the TE0 mode on the left side. It passes through two tapers that increase the waveguide width, making it multimode such that it supports TE0 and TE1 modes after the first taper and TE0, TE1 and TE2 modes after the second taper. The optical field intensity plots of modes supported in the waveguide after the second taper are illustrated in in Fig. 1(d). After passing through the two tapers, the input optical power remains predominantly in the TE0 mode, which gets modulated by the microwave signal applied to the travelling wave electrodes, similar to a conventional phase modulator. A schematic cross-section of a modulator in the SiN loaded thin-film lithium niobate on insulator platform is shown in Fig. 1(c). Afterwards, the multimode waveguide is tapered down in two stages back to the width of a single mode waveguide, which is then looped back to the input side, where it is coupled as the TE1 mode to the multimode waveguide, using mode multiplexers described in our previous work [18]. The TE1 mode passes again through the same traveling wave electrodes, experiencing a second modulation. Subsequently, the TE1 mode is coupled to a single TE0 mode waveguide using a mode multiplexer. The single mode waveguide is looped back again, where it is coupled as a TE2 mode to the multimode waveguide. The TE2 mode is then modulated again before it is coupled out as a TE0 mode to a single mode output waveguide. FDTD simulated operation of the TE0-TE1 and TE0-TE2 mode converters is shown in Fig. 1(b) [18]. In principle the number of loops can be increased arbitrarily, but it comes at a cost of device footprint and might require wider electrode gaps, as the width of the multimode waveguides needs to increase to accommodate the additional higher order modes.

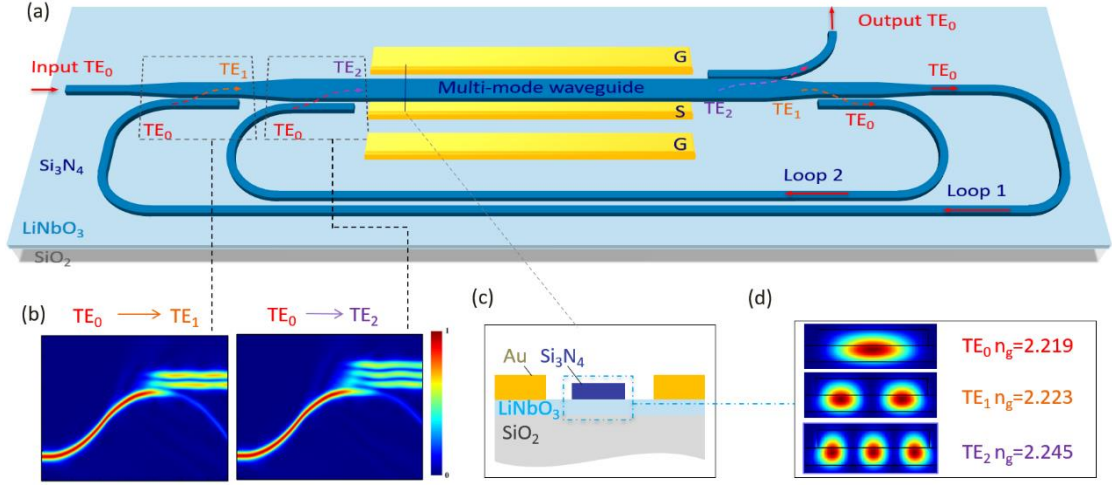

***Fig.1.*** *(a) Schematic of a recirculating phase modulator; (b) Simulated mode multiplexer for coupling the TE0 to TE1 mode and coupling the TE0 to TE2 mode; (c) Cross-section of the multimode phase modulator electrode region; (d) Normalized optical field plots for the TE0, TE1 and TE2 modes.*

### 3. Theoretical Analysis

In this section, we have a closer look at the recirculating modulator operation and what considerations need to be taken into account to achieve an increased modulation efficiency. When a sinusoidal signal $V(t) = A\sin(\omega_s t)$ is applied to a single pass phase modulator, the output optical field from the device can be expressed as:

$$E_{out} = E_o e^{i\omega_c t} e^{iR \sin \omega_s t} = E_o e^{i\omega_c t} \sum_{n=-\infty}^{n=+\infty} J_n(R) \, e^{in\omega_s t} \qquad (1)$$

where $E_o$ and $\omega_c$ are the amplitude and angular frequency of the optical carrier; $\omega_s$ is the angular frequency of the radio frequency (RF) modulating signal, $R$ is the modulation index, which is equal to $\pi A/V_\pi$, where $A$ is the amplitude of microwave signal and $V_\pi$ is the voltage required to shift the phase of the carrier by $\pi$, and $J_n$ is the Bessel function of the first kind. In the case of a recirculating modulator, as introduced in Fig. 1(a), the light is modulated multiple times, which for $N$ recirculations can be expressed as:

$$E_{out} = E_o e^{i\omega_c t} e^{iR_0 \sin \omega_s t} e^{iR_1 \sin \omega_s(t-\Delta T_1)} e^{iR_2 \sin \omega_s(t-\Delta T_2)} \ldots e^{iR_N \sin \omega_s(t-\Delta T_N)} \quad (2)$$

where $R_j$ ($j = 0, 1, 2, \ldots N$) is the modulation index of the $j$th recirculation and $\Delta T_j$ ($j = 1, 2, \ldots N$) is the loop transit time for light to re-enter the modulator, which is effectively the time delay between two different modes be modulated by the RF electrode. The loop transit time is given by $\Delta T_j = \Delta L_j / v_g$, where $\Delta L_j$ is the length of loop $j$ and $v_g$ is the group velocity of the fundamental mode TE0 in the loop waveguides. If one assumes that the modulation index is the same for all modes, equation 2 can be written as:

$$E_{out} = E_o e^{i\omega_c t} e^{i\, R * \kappa * \sin(\omega_s t - \varphi)} \qquad (3)$$

where $\varphi = \arctan \frac{\sin \omega_s \Delta T_1 + \sin \omega_s \Delta T_2 + \cdots + \sin \omega_s \Delta T_n}{1 + \cos \omega_s \Delta T_1 + \cos \omega_s \Delta T_2 + \cdots + \cos \omega_s \Delta T_n}$ and the $\kappa$ is the enhancement factor of the recirculating modulator. In the case of $N$th recirculation, the enhancement factor can be expressed as:

$$\kappa = \sqrt{(1 + \cos \omega_s \Delta T_1 + \cos \omega_s \Delta T_2 + \cdots + \cos \omega_s \Delta T_N)^2 + (\sin \omega_s \Delta T_1 + \sin \omega_s \Delta T_2 + \cdots + \sin \omega_s \Delta T_N)^2} \quad (4)$$

Then, expanding equation 3 by Bessel functions of the first kind, the output signal can be described as:

$$E_{out} = E_o e^{i\omega_c t} * \sum_{n=-\infty}^{n=+\infty} J_n(R * \kappa) * e^{in(\omega_s t - \varphi)} \quad (5)$$

As one can see from Eq. (4) and Eq. (5), the output signal is strongly dependent on the enhancement factor, which is determined by modulation frequency $\omega_s$ and the delay time $\Delta T_j$ for each recirculating loop. Maximum enhancement factor can be achieved when $\Delta T_j = m\pi/\omega_s$, where m is an integer. For $N$ recirculations, the maximum enhancement factor is equal to $N + 1$.

## 4. Device Design

For investigating the recirculating modulator, we chose the SiN loaded thin-film lithium niobate on insulator platform, see Fig. 1(a) and our pervious works [18, 19], with a 0.3 µm thick lithium niobate film and a 0.3 µm thick silicon nitride thickness. The waveguide widths needed to support the three different waveguide modes were 1.2, 2.8 and 4.5 µm in accordance with the mode multiplexers in [18]. For the travelling wave electrode, we chose an electrode separation of 9.5 µm and a gold metal thickness of 0.5 µm. The $V_\pi$ of the TE0, TE1 and TE2 modes was calculated to be 7.38, 7.35 and 7.52 V·cm.

In designing prototype recirculating phase modulators we chose a 25 GHz operation frequency. When considering the group index of the TE$_0$ in in loop waveguides $n_g = 2.23$, calculations show that for the device to operate at its highest enhancement factor the waveguide lengths for TE0 and TE1 loops have to respectively be 21.5 mm and 16.1 mm. To confirm this, we simulated the enhancement factor as a function of the modulation frequency, which is shown in Fig. 2. One can see that the conventional single-pass phase modulator has an enhancement factor of 1 for all frequencies (assuming ideal operation). Conversely, a single loop recirculating modulator has a periodic response, reaching an enhancement factor of $\kappa$~2 at 0, 6.3, 12.5, 18.8, 25, 31.3 and 37.5 GHz, while the dual loop recirculating modulator reaches an enhancement factor of $\kappa$~3 for 0 and 25 GHz. This shows that we should expect an increase of the enhancement factor to 3 when looping the light back twice and operating at 25 GHz with delay lengths of 21.5 mm and 16.1 mm for the two loops.

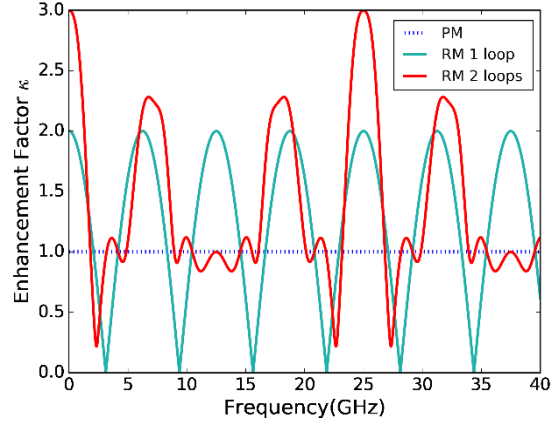

*Fig. 2. Enhancement factors as a function of the modulation frequency for single and a dual loop recirculating phase modulators A typical single-pass phase modulator exhibits a frequency independent baseline κ = 1 enhancement factor.*

## 5. Experimental Results

To prove the device performance, we fabricated the designed 25 GHz operation frequency recirculating modulator devices with one and two loops and benchmarked them against a traditional single-pass phase modulator. Fabrication was performed following the steps outlined in our prior work [18]. In brief, optical waveguiding in thin-film lithium niobate (see Fig. 3(b)) is established by depositing a SiN loading layer by means of reactive sputtering [20], followed by patterning this SiN film using electron beam lithography and reactive ion etching. Gold electrodes (Fig. 3(c)) comprising the electrooptical modulator section were formed using laser lithography patterning with a subsequent physical vapor deposition and lift-off process. The fabricated device with two recirculation loops is shown in Fig. 3(a). The on-chip footprint f this device is 10.85 mm × 5.61 mm and the active modulation section takes up 62% of total device length.

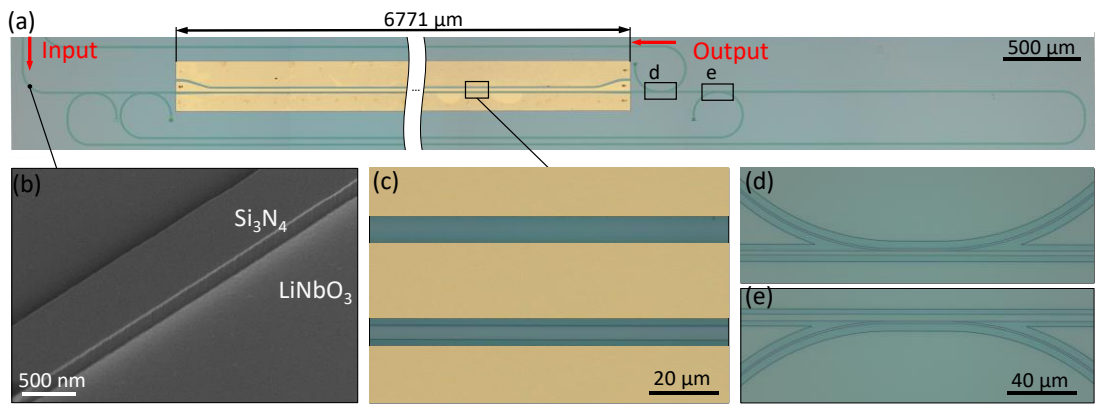

*Fig. 3 (a) Optical microscope image of recirculating phase modulator device with two loops; (b) Scanning electron micrograph of a fabricated single mode waveguide; (c) shows the traveling wave electrode alignment to the multimode waveguide; (d) shows the fabricated mode multiplexer to couple the TE2 to a TE0 mode and (e) is the mode multiplexer for coupling the TE1 mode to a TE0 mode.*

To characterize the performance of the recirculating modulators, we couple 1550 nm light into the devices and applied an amplified 25 GHz microwave signal from an Anritsu MG3694A frequency synthesizer to the traveling wave electrodes and measured the resulting spectra with a Finisar WaveAnalyzer 1500S High-Resolution Optical Spectrum Analyzer. The output of the electrodes was terminated with 50 Ohm impedance matching terminators. Fig. 4(a-c) shows the measured spectra when a 25 dBm RF power is applied to a typical single-pass phase modulator as well as recirculating phase modulators with one and two loops, respectively. These modulators have the same electrode length, gap, and thickness. One can see that the overall optical power levels for the graphs are slightly different, which is most likely caused by fabrication errors as well as increased losses for the recirculating modulators with two loops, due to the increased waveguide length and additional mode multiplexing structures. One can also see that a recirculating modulator can generate more sidebands compared to a conventional phase modulator. To estimate the modulation index, we fitted the power of sideband power using Eq. (1) and found that the modulation index approximately doubled for the single loop recirculating modulator and tripled for the two loop recirculating modulator (see black dots in Fig. 4(a-c)), when compared to a single-pass phase modulator. These results matched well with our prediction of how the enhancement factor scales with the number of loops, as shown in Fig. 2.

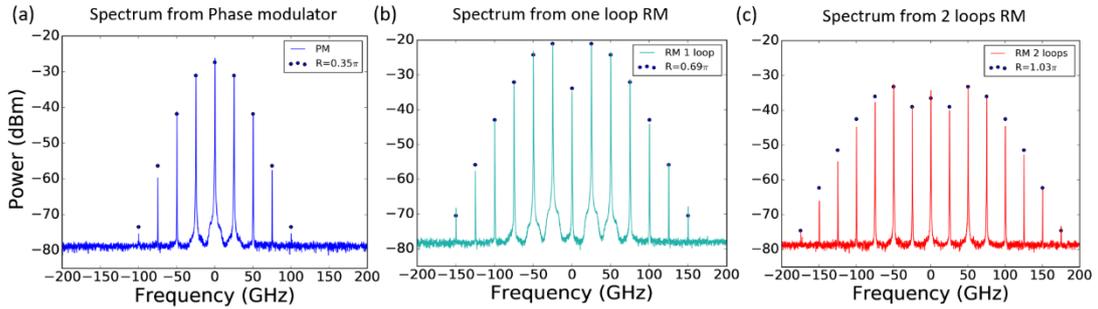

*Fig. 4. Measured spectra for (a) single mode phase modulator, (b) single loop and (c) two loop recirculating modulators when 25dBm RF power is applied.*

Next, we characterized the modulation efficiency of the different modulators by adjusting the applied RF power so that the spectra exhibit the same modulation index (same power distribution of the sidebands) as compared to the single-pass phase modulator. The results are shown in Fig. 5 (a-c), which show that the single-pass phase modulator as well as the single loop and dual loop recirculating phase modulators exhibit the same modulation index, while the applied microwave power decreases from 28 dBm, to 22.1 dBm and 17.7 dBm, respectively. This corresponds well to the theoretical prediction of the RF power reduction for a single loop recirculation, which is $2^2$ (6 dB), and for a dual loop recirculation is $3^2$ (9.5 dB). This indicates that the dual loop recirculating phase modulator requires approximately one order of magnitude less RF power to achieve the same modulation index, increasing the modulation efficiency significantly.

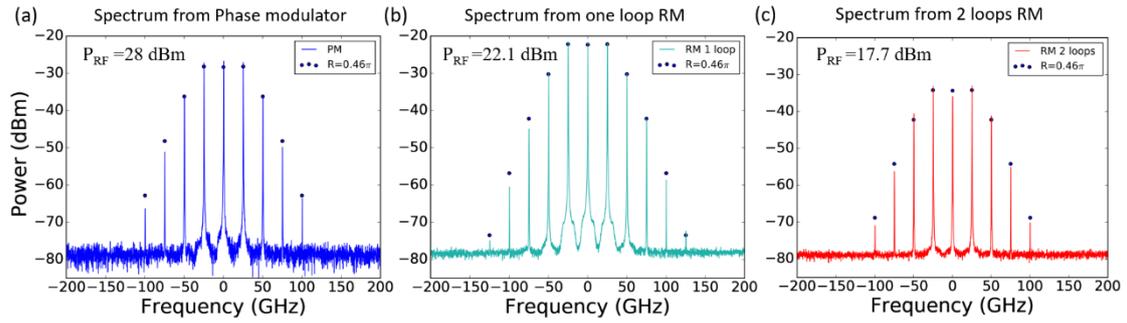

*Fig. 5. Equivalent spectra generated by (a) single mode phase modulator (28 dBm RF power), (b) one loop (22.1 dBm RF power), and (c) two loop recirculating phase modulators (17.7 dBm RF power).*

To analyze the modulation efficiency, we measured the modulation index of the modulators for different RF powers. The results are presented in Fig. 6. One can see that the modulation index has a near quadratic dependence on the RF power, which matches the RF power scaling with applied voltage. One can also observe how the modulation index for the recirculating modulators increases by a factor of ~2 for a single loop recirculating modulator and ~3 for a dual loop recirculating modulator compared to the single pass phase modulator with an equivalent electrode configuration. This demonstrates that light recirculation in modulators can notably improve their modulation efficiency.

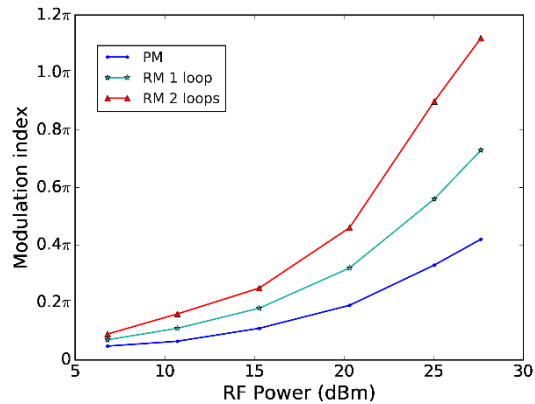

*Fig. 6. Modulation index of a conventional phase modulator and the recirculating modulators with one and two loops as a function of the applied RF power.*

## 6. Discussion

Experimental demonstration of the recirculating modulator has shown that increasing the number of loops does indeed increase the modulation efficiency. Although this is highly attractive, one also needs to consider trade-offs that come with such a design. For example, one should be aware that the operation bandwidth of the modulator is decreased and that adding more loops to the recirculating phase modulators may result in higher optical losses. There is also a practical limitation on the loop number scalability, as higher order multimode waveguides will be required, causing the waveguide modes to be closer in their effective indices, which can lead to enhanced cross talk between the modes. Furthermore, a wider multimode waveguide will increase the separation of electrodes, resulting in a higher $V_\pi$ of modulators. Nevertheless, we believe that the investigated recirculating phase modulators are highly attractive for applications for the generation of electro optical frequency combs [21, 22], where one can apply a known and constant RF signal. The advantage of the recirculating modulator is that RF losses of the electrodes are less of a concern compared to high efficiency modulators that rely on long electrodes. It may also become possible to consider such devices for optical computational tasks, as the looping back through the same modulator offers the imprinting of different electrical signals in time onto the optical domain [23]

## 7. Conclusion

In this work, we present a new recirculating phase modulator concept, explore its properties through numerical simulations and experimentally validate the performance of a prototype device fabricated on the thin-film lithium niobate on insulator platform. The recirculating modulators with two light recycling loops exhibit an approximately threefold increase in modulation efficiency, enabling a reduction of the microwave power required for achieving the same modulation index by approximately one order of magnitude. We also provide the numerical expression of enhancement factor for $N$th recirculation case, which provides the possibility for further improvement in the future. Moreover, we believe such modulators pave the way to reduce the power required for electro-optical frequency comb generation and for executing optical computation tasks.

## 8. Funding

The material is based on work supported by the Australian Research Council (ARC) under award ARC DP190102773.

## 9. Data availability

The data that support the findings of this study are available from the corresponding author upon reasonable request.